\documentclass[onecolumn]{aastex63}
\usepackage{graphics,amsmath,amsbsy}

\newcommand{\lapprox} {\, \lower3pt\hbox{$\sim$}\llap{\raise2pt\hbox{$<$}}\,}
\newcommand{\gapprox} {\, \lower3pt\hbox{$\sim$}\llap{\raise2pt\hbox{$>$}}\,}

\newcommand{\ang}{\mathrm{\AA}}

\definecolor{mrkred}{RGB}{0,0,0}
\newcommand{\mrkred}[1]{{\color{mrkred}  #1}}

\definecolor{mrk}{RGB}{0,0,0}

\renewcommand{\vec}[1]{ \protect {{\mathbf{\boldsymbol{#1}}}}}

\shorttitle{Particle acceleration and escape into the heliosphere}
\shortauthors{Gordovskyy et al.}
\begin{document}
\title{Particle acceleration and their escape into the heliosphere in solar flares with open magnetic field}

\author[0000-0003-2291-4922]{Mykola Gordovskyy}
\affiliation{Department of Physics, Astronomy \& Mathematics, University of Hertfordshire, Hatfield AL10 9AB, UK}
\affiliation{Department of Physics \& Astronomy, University of Manchester, Manchester M13 9PL, UK}

\author[0000-0002-7089-5562]{Philippa K. Browning}
\affil{Department of Physics \& Astronomy, University of Manchester, Manchester M13 9PL, UK} 

\author[0000-0002-6814-6810]{Kanya Kusano}
\affil{Institute for Space-Earth Environmental Research, University of Nagoya, Nagoya 464-8601, Japan}

\author[0000-0002-6814-6810]{Satoshi Inoue}
\affil{Center for Solar-Terrestrial Research, New Jersey Institute of Technology, Newark, NJ 07102-1982, USA}

\author[0000-0002-7089-5562]{\fbox{Grigory E. Vekstein}}
\affil{Department of Physics \& Astronomy, University of Manchester, Manchester M13 9PL, UK}

\begin{abstract}
Energetic particle populations in the solar corona and in the heliosphere appear to have different characteristics even when produced in the same solar flare. It is not clear what causes this difference: properties of the acceleration region, the large-scale magnetic field configuration in the flare, or particle transport effects, such as scattering. In this study we use a combination of magnetohydrodynamic and test-particle approaches to investigate magnetic reconnection, particle acceleration and transport in two solar flares: an M-class flare on June 19th, 2013, and an X-class flare on September 6th, 2011. We show that in both events , the same regions are responsible for the acceleration of particles remaining in the coronal and being ejected towards the heliosphere. However, the magnetic field structure around the acceleration region acts as a filter, resulting in different characteristics (such as energy spectra) acquired by these two populations. We argue that this effect is an intrinsic property of particle acceleration in the current layers created by the interchange reconnection and, therefore, may be ubiquitous, particularly, in non-eruptive solar flares with substantial particle emission into the heliosphere. 
\end{abstract}

\section{Introduction}
\label{s-intro}

Observations in X-ray, microwave and radio domains demonstrate that high-energy electrons and ions, propagating at significant fractions of the speed of light, carry most of the energy released in flares \citep{hole11,kone11,kone19}. The majority of these energetic particles precipitate towards the photosphere. Energetic electrons in the lower solar atmosphere produce bright microwave and hard X-ray (HXR) emissions, heating the corona and chromosphere. However, some of the energetic particles escape into the heliosphere, becoming observable first via deca-metric radio emission in the upper corona, then directly in the heliosphere and near the Earth by spacecraft. These particles are an essential component of space weather and can either cause, or serve as an early warning of, problems with satellite communication and navigation systems, power grids and avionics \cite[e.g.][]{pose09}. Hence, it is important to understand how energetic particles are produced and transported in the flaring corona and ejected into the heliosphere.

Several previous studies have compared observed properties of energetic particle populations in the corona and the heliosphere. It appears that these properties are different even when it is clear that the both populations are produced by the same solar flares \citep[e.g.][]{krue07,klda17}. They demonstrate that, at least after short-duration solar flares, the properties of the two particle populations correlate but are clearly different. The question is what causes this difference: is it because the particles precipitating in the corona and ejected into the heliosphere are accelerated at different locations or by different mechanisms in the flaring corona, or because these populations of particles change differently during transport?

In principle, both acceleration and transport effects can play an important role. Since solar flares are very varied in terms of the dynamics, energetics and magnetic field configuration, it is practically impossible to create a universal phenomenological model and establish the relative effects of the mechanisms on the formation of observed properties of different particle populations universal for the majority of flares. Instead, to investigate what mechanisms affect particle transport and lead to the difference between the electron populations going into the heliosphere and precipitating in the corona, comprehensive modelling of magnetic field, thermal and non-thermal plasmas in realistic, individual flare configurations is needed.   

\citet{gore20} created a realistic MHD-test-particle (MHDTP) model of an actual solar flare, successfully matching the observations, demonstrating that MHDTP is a viable approach for modelling aimed at reproduction of magnetic field and non-thermal plasma behaviour in actual flares. In this study the MHDTP approach is used to investigate particle acceleration in two flares, with the goal to model  and explain the properties of populations of both precipitating and escaping particles. We consider the behaviour of the magnetic field and hot plasmas (Section~\ref{s-models-mhd}), investigate the dynamics of energetic particles (Section~\ref{s-models-particles}), derived synthetic observables and compare them with actual observations (Section~\ref{s-models-obs}). The main outcomes of this study, linking the energetic particle properties with the characteristics of the energy release regions, are discussed in Section~\ref{s-summary}.

\section{Main equations and numerical set-up}
\label{s-models-setup}

This study uses the approach developed by \citet{gore14} and validated for individual data-driven flare modelling by \citet{gore20}: it combines 3D time-dependent MHD modelling with test-particle simulations using the guiding centre approximation.

The MHD model of each considered solar flare is initiated using the magnetic field configuration assuming that the magnetic field before the flares is in non-linear force-free (NLFF) state. NLFF reconstructions are based on photospheric vector magnetograms obtained by SDO/HMI.  The initial magnetic field distributions are calculated using the approach and numerical code developed by \citet{inoe14}. The calculations are carried out on a Cartesian grid with grid-points uniformly spaced in each direction. For the two considered events -- the 06 September 2011 flare  (Flare~1, thereafter) and the 19 June 2013 flare  (Flare~2, thereafter) the grid has 512$\times$256$\times$256 elements, representing a volume with the dimensions of 354.4$\times$177.2$\times$177.2~Mm. Initial pressure and density distributions in each model are assumed to be gravitationally-stratified, plane-parallel, formed by two regions: 2~Mm-wide region with the temperature of 0.01~MK, representing the chromosphere, below $\sim$85~Mm-wide volume with the temperature of 0.9~MK, representing quiet corona. Plasma velocity is assumed to be zero everywhere in the model domains at the start of MHD simulations.

The evolution of the magnetic field and thermal plasma is then followed using single-fluid 3D resistive MHD simulations performed with the Lare3D code \citep{arbe01}. The MHD simulations are carried out using the same Cartesian grid as in the NLFF reconstructions. The resistivity is assumed to be zero at the bottom boundary, {\it i.e.} the magnetic field is ``frozen-in''. The velocities at the lower boundary of the models are set to be zero, effectively, meaning the normal magnetic field at this boundary does not change. In the rest of the numerical domains, the resistivity depends on the local values of the electric current density, plasma temperature and density: 
\begin{eqnarray}
\label{eq-eta}
&&\eta = \eta_0+\eta_1,\\
&&\eta_0 = 10^{-5} \nonumber\\
&&\eta_1 = \begin{cases}
0, & \; j < j_\mathrm{crit}\\
10^{-3}, & \; j \geq j_\mathrm{crit}. \nonumber
\end{cases}
\end{eqnarray}
Here, $j_\mathrm{crit}$ is the critical current, which is the threshold for the anomalous resistivity, described in more detail below. For the sake of convenience, in this manuscript we use magnetic diffusivity to quantify resistivity, i.e. $\eta=\frac 1{\mu_0 \sigma}$ and, unless otherwise stated, it is 
measured in units of 8$\times$10$^{12}$~m$^2$~s$^{-1}$. \mrkred{The critical current $j_\mathrm{crit}$ depends on the local density and temperature to mimic anomalous resistivity triggered by ion-cyclotron instability. Thus, the anomalous resistivity is triggered when the current drift velocity exceeds local proton thermal velocity, i.e. $j>env_{p.th}$ \citep{bare11a,bare11b,gore14}. For typical conditions in the pre-flaring corona ($n= 10^{15}$~m$^{-3}$ and $T=1$~MK) the value of critical current will be order of 10~A~m$^{-2}$.  However, it is important to note, that the width of the current layer in large-scale MHD models is grossly overestimated: while the real width is expected to be equal to several proton Larmor radii, i.e. for $B\sim 0.001$T and $T\sim 10$MK (temperature in the flaring corona) it should be around $\Delta\sim$1--10~m, in MHD models it cannot be smaller than the grid step. Hence, in our models with the grid step of $\delta L \approx 7\times 10^5$~m, RCS thickness is overestimated by factor of $\sim 10^5$, which, in turn, means that the current density in RCS is underestimated by the same factor. That is why we introduce the scaling parameter $\mathcal{K} = L_R/\delta L = 10^{-5}$ to account for this effect and make $j_\mathrm{crit}$ comparable to the maximum current values in the initial configurations of both models. For the typical coronal values in our models (n$\sim$10$^{15}$~m$^{-3}$ and T$\approx$0.9~MK) this brings the value of the critical current to about 1.5$\times$10$^{-4}$~A~m$^{-2}$. The value of magnetic diffusivity corresponding to anomalous resistivity $\eta_1$ in Equation~\ref{eq-eta} is tuned to provide realistic event duration in the model; more specifically, so that the durations of the fast energy release phases in the models are comparable to the durations of the flash phase in the actual events. The adopted value of $\eta_1$ corresponds to 8$\times$10$^9$~m$^2$~s$^{-1}$, i.e. a feature with the size of $\delta L \approx 7\times 10^5$~m dissipates within $\sim 100$~s.}

The MHD models are then used to trace trajectories of large numbers ($\sim$10$^6$) of test electrons and protons using relativistic guiding-centre motion equations with the field values calculated for each particle position at each time step using four-linear interpolation (in the $[X,Y,Z,t]$ space). These calculations are performed using the 3D GCA code \citep{gobr11,gore14,gore19}.

The initial test-particle positions in phase space replicate thermal plasma distributions at the start of the MHD simulations, i.e. they have the Maxwellian distributions in respect of the absolute velocity with the temperature, characterising the distribution being approximately corresponding to the local plasma temperature in the MHD model. The spatial distributions of the test-particles approximately replicate the density distribution at the beginning of MHD simulations. This is done partially by using different statistical weights of test-particles. This makes the initial test-particle populations statistically representative of the thermal plasma at the beginning of MHD simulations.

'Thermal bath' conditions are applied to test-particles at all six boundaries. This means that for every test-particle leaving the domain, a particle  of the same species in injected into the domain at the same location. The pitch-angle of the injected particle is chosen randomly from the isotropic distribution (although, effectively, it means that only half of the pitch-angle space, as only particles travelling inward can be selected). The absolute velocity of the injected particle is chosen randomly from the Maxwellian distribution with the temperature corresponding to the local temperature at the beginning of the corresponding MHD simulations.

The behaviour of all test-particles is observed using fixed-order lists, which include 'active' particles (i.e. those still remaining in the computational domain) and 'left' particles. 

\section{Solar flare models}
\label{s-models}

\subsection{Magnetic reconnection and energy release}
\label{s-models-mhd}

Let us now discuss the  magnetic field and plasma behaviour in the two considered events. 

\begin{figure}
\centering{\includegraphics[width=0.9\textwidth]{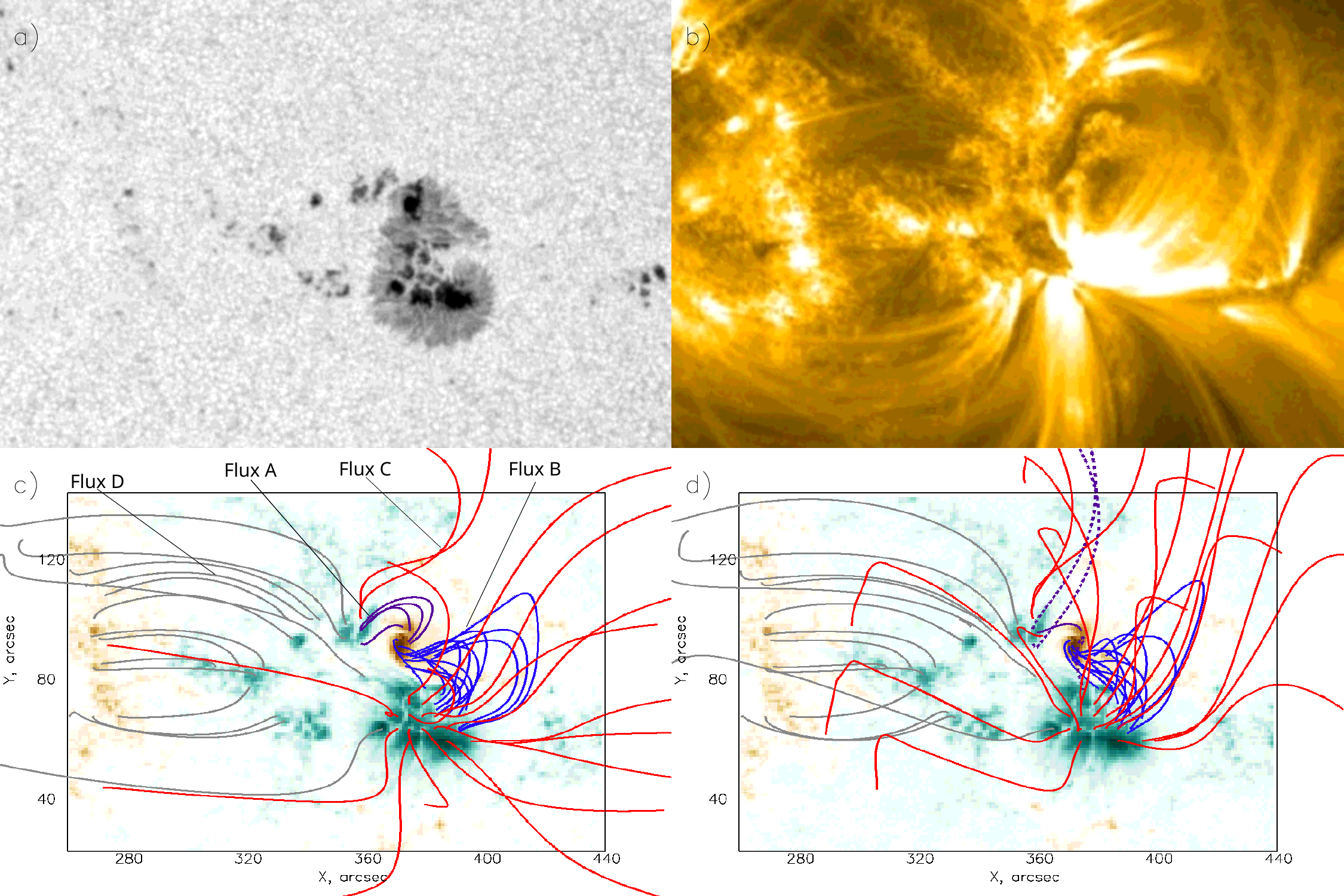}}
\centering{\includegraphics[width=0.4\textwidth]{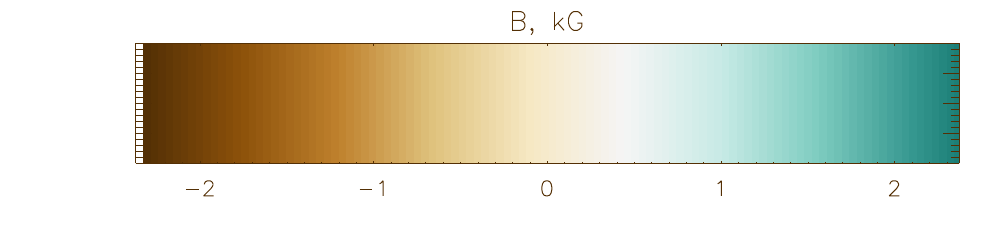}}
\caption{White-light (panel a) and 131$\mathrm{\mathring{A}}$ extreme ultra-violet (EUV, panel b) intensity maps of the active region 11283 observed about one hour before Flare~1 on 6 September 2011 by SDO/HMI and SDO/AIA, respectively. Panels (c) and (d) show selected magnetic field lines from the MHD simulations over the magnetograms observed by SDO/HMI at the beginning and the end of simulations, respectively. Different \mrkred{line colours denote different magnetic connectivities: flux A (purple), flux B (blue), flux C (red), and flux D (grey) (see text for details)}. Dashed field lines in panel (d) are those that changed connectivity.}
\label{f-flare1ini}
\end{figure}

The first event, an X-flare, which occurred on September 6th, 2011 was observed over an active region AR11283 at around 21:45~UT. The initial configuration in Flare~1 is shown in Figure~\ref{f-flare1ini}(a-c). The optical continuum (Figure~\ref{f-flare1ini}a) and the magnetogram (Figure~1c) show that the coronal magnetic field structure in this active region is defined by a close group of two large strong asymmetric sunspots with magnetic field strengths of about 2.0--2.2~kG, and diffuse photospheric network field surrounding them. The positive flux through the photosphere in this region appears to be insufficient to fully compensate the negative flux and, hence, some of the magnetic field in this regions is open.

In terms of magnetic connectivity, it is possible to distinguish four separate structures. The first structure is the magnetic field connecting the patch of negative polarity north-east of the two sunspots with the positive sunspots (Flux~A, purple lines in Figure~\ref{f-flare1ini}c). The second structure (Flux~B, blue field lines  in Figure~\ref{f-flare1ini}c) is the field connecting the large negative and positive sunspots. Some of this field is noticeably twisted, indicating that it may store a significant amount of free energy. The third structure is the open magnetic field originating mainly from the patch of negative polarity corresponding to a large sunspot, as well as from the diffuse magnetic field of negative polarity surrounding positive sunspot (Flux~C, red lines). (It should be noted that in this paper by open magnetic field we mean the field connecting the lower and upper boundaries of the computational domain.). Finally, there is a flux connecting diffuse area of negative polarity with a diffuse area of positive polarity in the eastern (left) part of the considered region (grey lines, Flux~D). This relatively weak field does not reveal any significant evolution during the event

\begin{figure}
\centering{\includegraphics[angle=90,origin=c, width=0.5\textwidth]{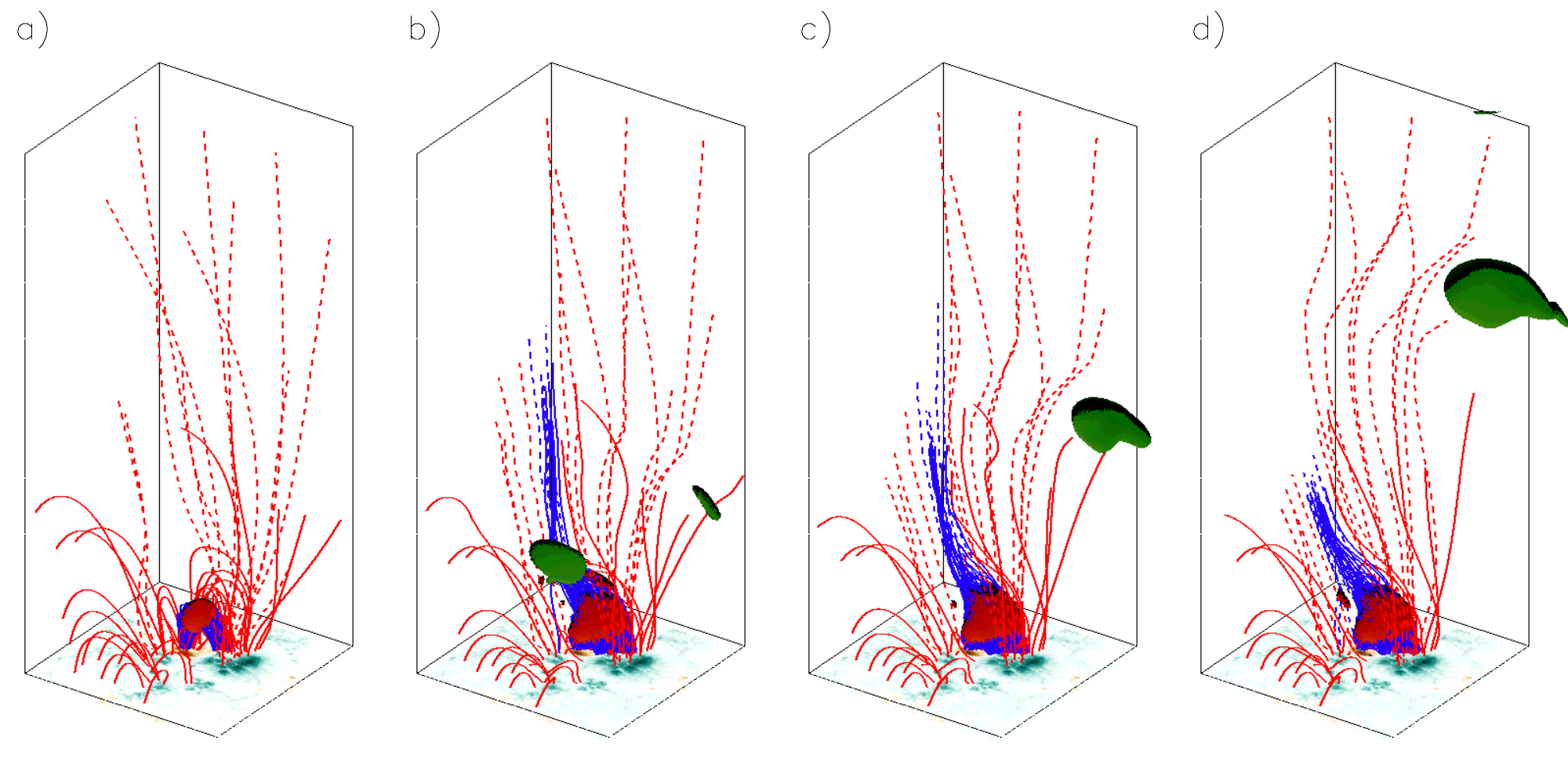}}
\caption{Selected magnetic field lines in Flare~1 at different times: t=0 (panel a), t=72s (panel b), t=152s (panel c), t=508s (panel d). \mrkred{Dark red surface denotes the location of the diffusion region, i.e. the volume where high local current density triggers anomalous resistivity, resulting in high electric field.} Blue lines are magnetic field lines penetrating the diffusion region. Dashed lines connect lower and upper boundaries of the domain. Green surface shows the volume where plasma velocity is 0.75 of the current maximum velocity in the domain, which corresponds to 0.18, 0.13, 0.11 and 0.07~Mm/s in panels (a), (b), (c) and (d), respectively.}
\label{f-flare1mhd}
\end{figure}

\begin{figure}
\centering{\includegraphics[width=0.5\textwidth]{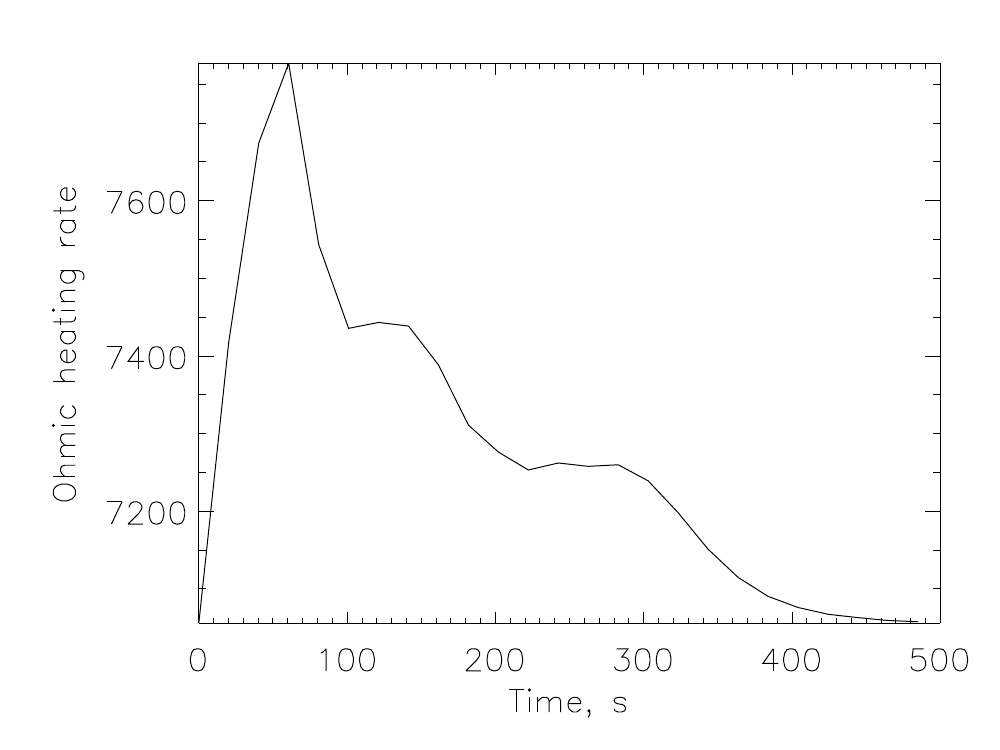}}
\caption{Total Ohmic dissipation (i.e. $\eta j^2$ integrated over the domain volume) in Flare~1.}
\label{f-flare1nrgj}
\end{figure}

\begin{figure}
\centering{\includegraphics[width=0.5\textwidth]{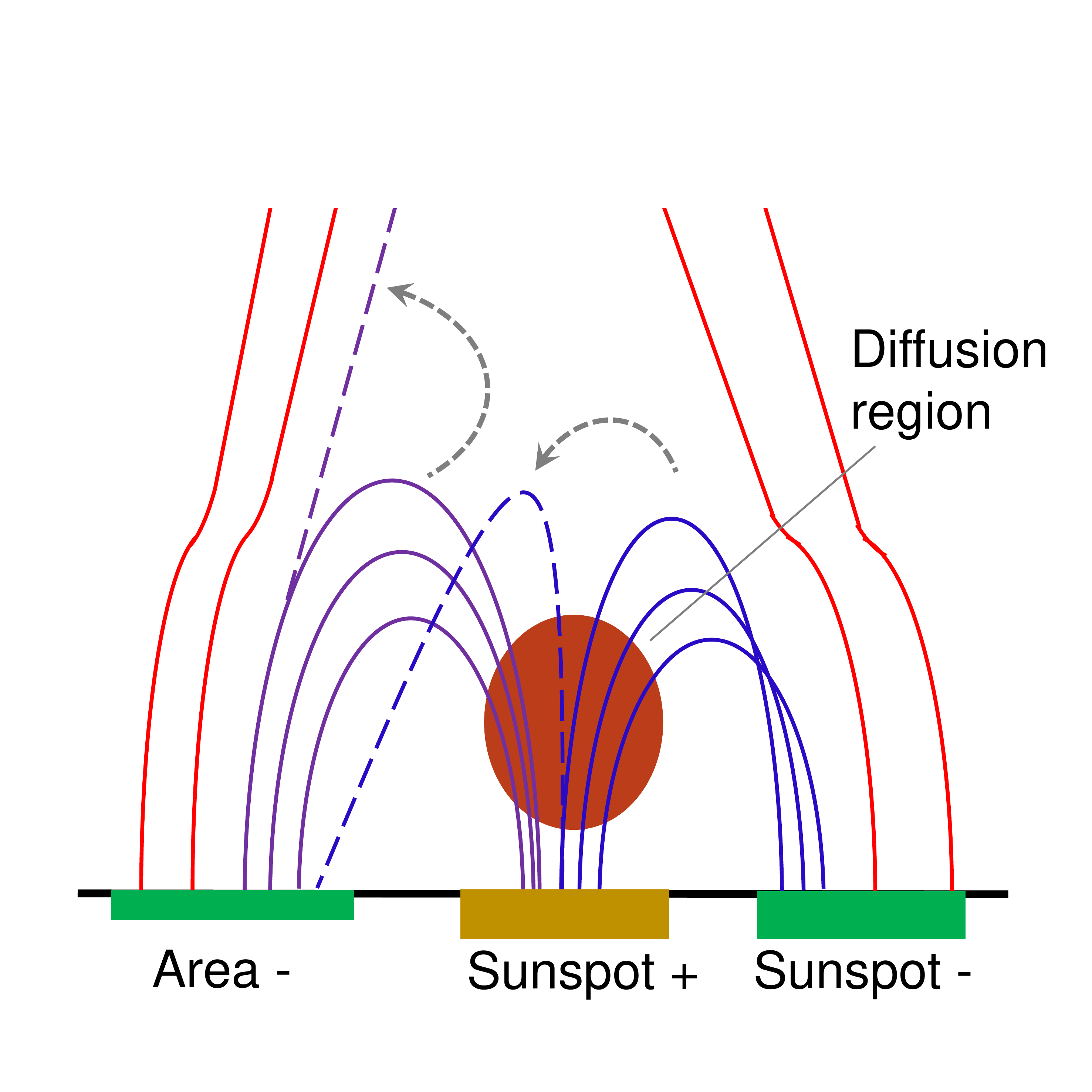}}
\caption{Sketch demonstrating the change of magnetic connectivity in Flare~1.}
\label{f-flare1sketch}
\end{figure}

Figure~\ref{f-flare1mhd} shows evolution of magnetic field and parallel electric field, and reveals locations of fast plasma flows during the impulsive phase (i.e. fast energy release stage) in the considered flare based on the MHD simulations described above. The diffusion region (dark red surface) denotes the volume with non-zero anomalous resistivity, i.e. the volume with fast magnetic energy conversion and intense particle acceleration due to high parallel electric field. Additionally, the configuration formed towards the end of the impulsive phase is shown in Figure~\ref{f-flare1ini}d. It can be seen that initially the volume containing high parallel electric field is located about 10~Mm above the negative sunspot and involves field from Fluxes~A and B (Figure~\ref{f-flare1mhd}a). Ohmic heating and resulting flows in this volume result in local magnetic field redistribution and geometrical expansion of the diffusion region. However, it should be noted that the Ohmic dissipation rate integrated over the flaring volume steadily decreases from about 15~s after the start of reconnection (i.e. after the start of model evolution, Figure~\ref{f-flare1nrgj}). As the reconnection and current dissipation progress, there is some change of connectivity. Firstly, some magnetic field originally forming flux~A becomes open (purple dashed lines in Figure~1d). Furthermore some magnetic field from flux~A connects to the large negative sunspot, becoming flux~B. The scheme of connectivity change is shown in Figure~\ref{f-flare1sketch}.

After the beginning of the impulsive phase, the fastest plasma flow can be seen just outside of the diffusion region (i.e. volume with high E$_{||}$), where reconnection and energy release occur. However, as the reconnection progresses, there appears an upward moving feature. It seems to originate at about 20--30~Mm approximately above the large sunspot of negative polarity and then moves up slightly in the southern direction with a speed of about 0.2~Mm/s and decelerates.
In terms of its origin, the fast upflow is likely to be similar to the reconnection plasma jets. Interestingly, this upflow appears to be organised in a series of nearly horizontal layers of fast plasma moving up. This leads to the 'kinks' in the shapes of magnetic field lines in Figure~\ref{f-flare1mhd}. Although of  unusual structure, we believe this upflow represents a real physical effect and is not a computational artefact, because (a) this effect is not observed in the model of Flare~2, and (b) it results in an observational feature very similar to one observed in Flare~1 (see Section~\ref{s-models-obs}).

\begin{figure}
\centering{\includegraphics[width=0.9\textwidth]{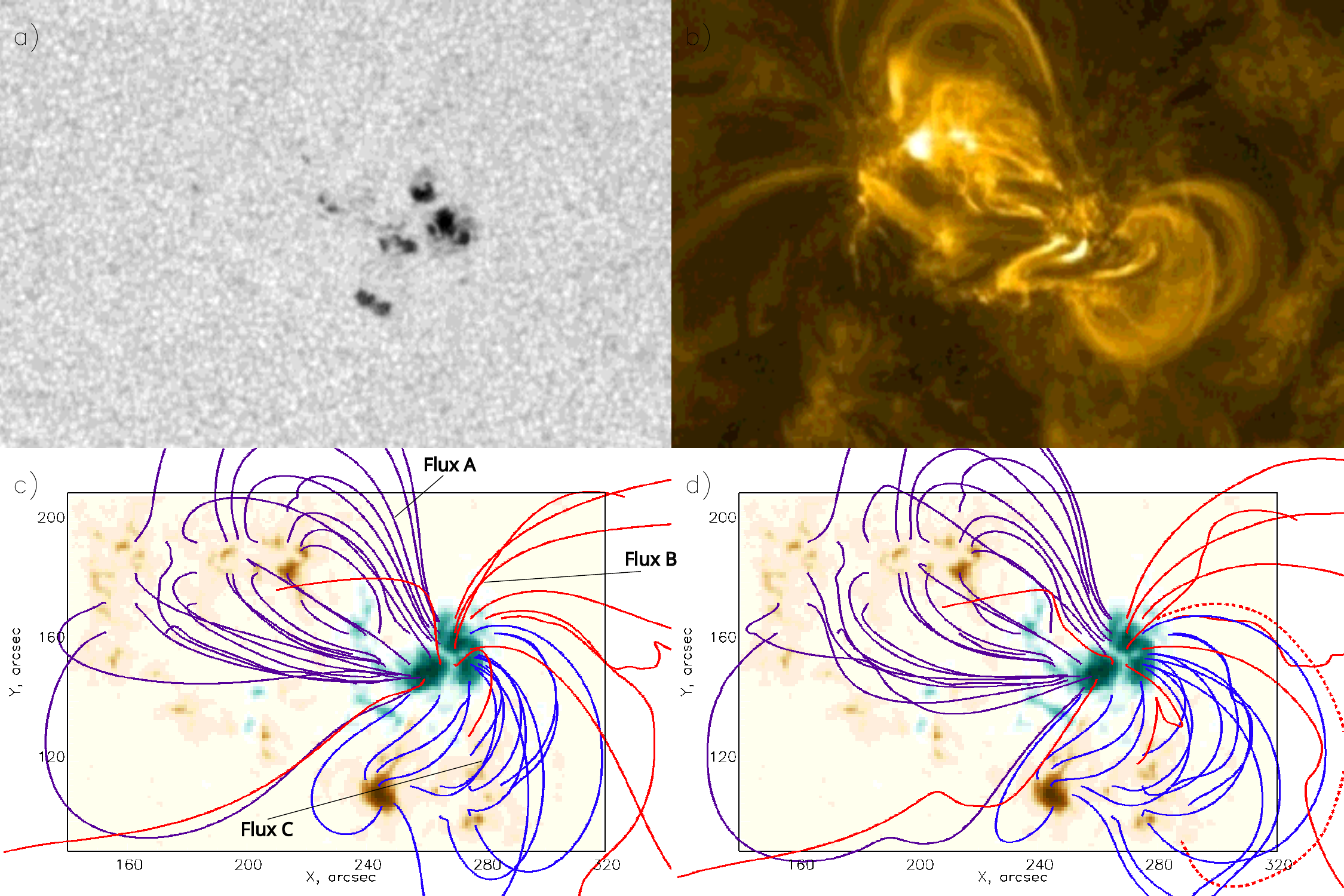}}
\centering{\includegraphics[width=0.4\textwidth]{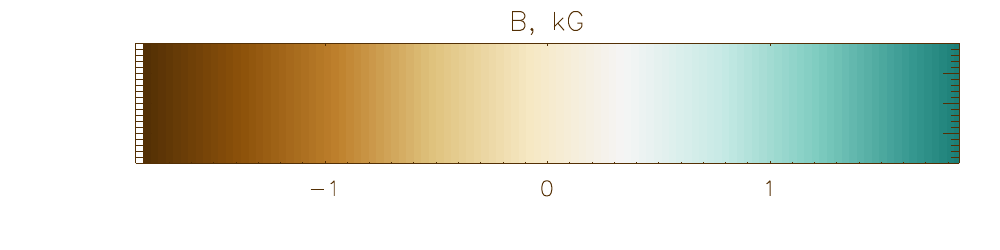}}
\caption{Same as in Figure~1 but for Flare~2. \mrkred{Different line colours denote different connectivities: flux A (purple), flux B (red), and  flux C (blue) (see text for details)}.}
\label{f-flare3ini}
\end{figure}

The second considered event was an M-class flare observed on June 19th, 2013, after around 07:00~UT in the active region 11776. Similar to the first event, it  occurred in an active region formed over a compact group of two intense sunspots with magnetic field of about 1.7--1.9~kG (Figure~\ref{f-flare3ini} a-c). This group also had non-zero net magnetic flux through the photosphere, and, hence, some open field.

The initial magnetic configuration for this event (Figure~\ref{f-flare3ini}c) can be split into three structures. The first one is the magnetic field connecting a large sunspot of negative polarity in the centre of active region with the diffuse positive area at the north-east (purple lines, Flux~A). The second structure in the open field originating in the negative sunspot in the centre of the group (red lines, Flux~B). The third structure is the magnetic field linking the two large sunspots (blue lines, Flux~C).

\begin{figure}
\centering{\includegraphics[angle=90,origin=c, width=0.5\textwidth]{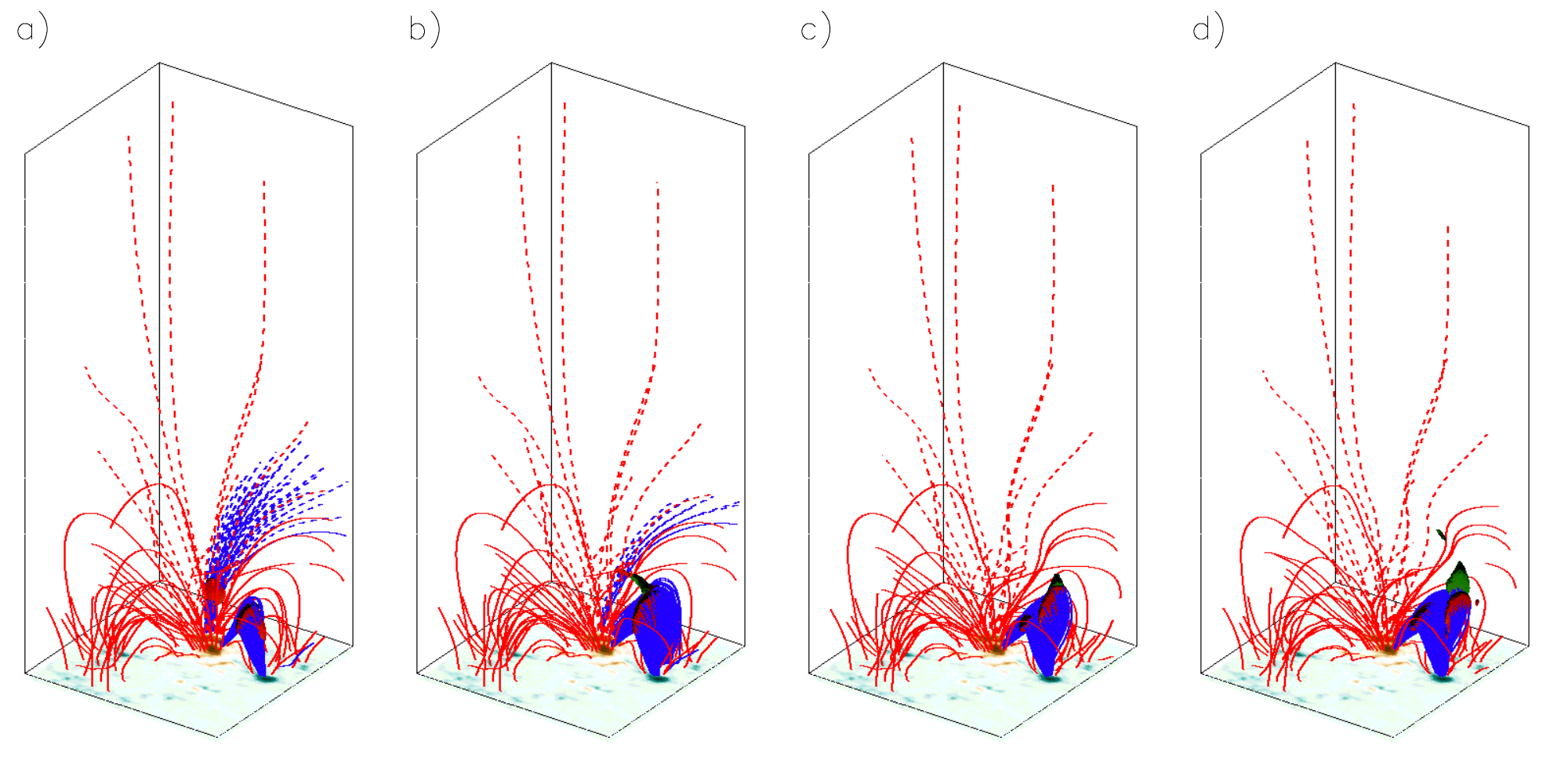}}
\caption{Selected magnetic field lines in Flare~2 at different times: t=0 (panel a), t=80s (panel b), t=164s (panel c), t=488s (panel d). Blue lines are magnetic field lines penetrating the magnetic diffusion region, which is shown as a dark red surface. Dashed lines connect lower and upper boundaries of the domain. Green surface shows the volume where plasma velocity is 0.75 of the current maximum velocity in the domain, which corresponds to 0.09, 0.06, 0.04 and 0.01~Mm/s in panels (a), (b), (c) and (d), respectively.}
\label{f-flare3mhd}
\end{figure}

\begin{figure}
\centering{\includegraphics[width=0.5\textwidth]{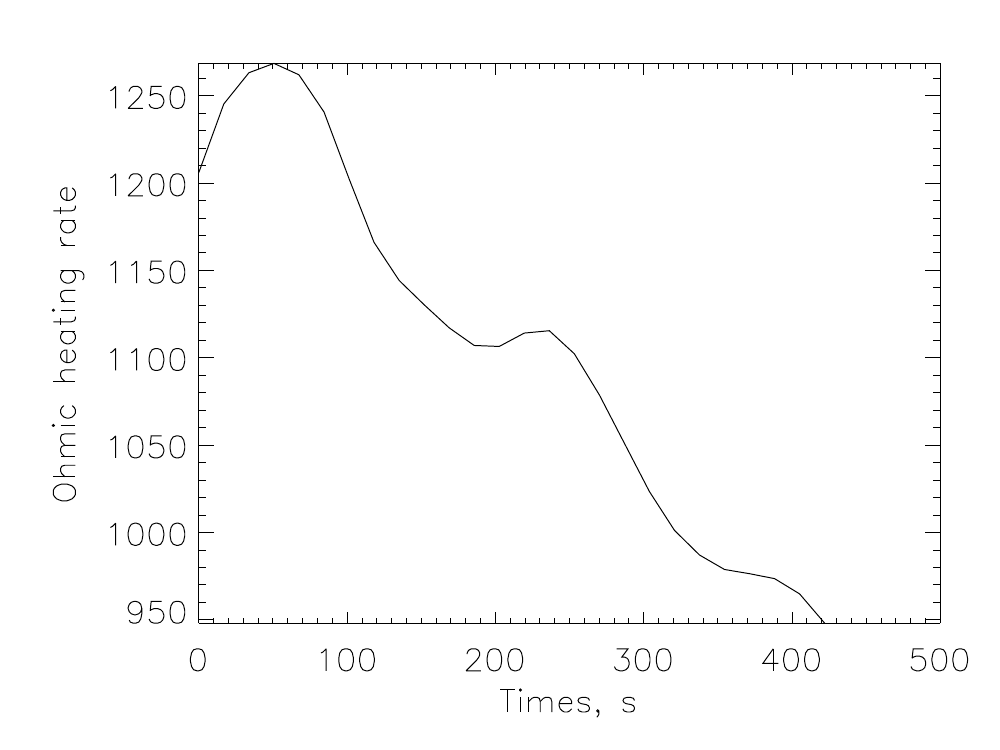}}
\caption{Same as in Figure~3 but for Flare~2.}
\label{f-flare3nrgj}
\end{figure}

The dynamics of the magnetic field is shown in Figure~\ref{f-flare3mhd}. It can be seen that at the beginning of the evolution the region with high parallel electric field (diffusion region) consists of two compact volumes: one sitting  mainly in the open magnetic field around 15~Mm over the negative sunspot, another, slightly larger, located within the flux~C, connecting large sunspots in the group. Similar to the model of Flare~1, the volume occupied by high parallel electric field also expands at the onset of simulations. However, the total Ohmic dissipation rate (a proxy of volume-integrated $j^2$) clearly decreases with time (Figure~\ref{f-flare3nrgj}). Interestingly, the Ohmic heating rate for the model of Flare~2 reveals some oscillatory behaviour, possibly indicating some slow MHD pulsations in the diffusion region \cite[see also][for recent results and discussion of oscillatory reconnection]{smie22,stee22,kare22}.

\begin{figure}
\centering{\includegraphics[width=0.5\textwidth]{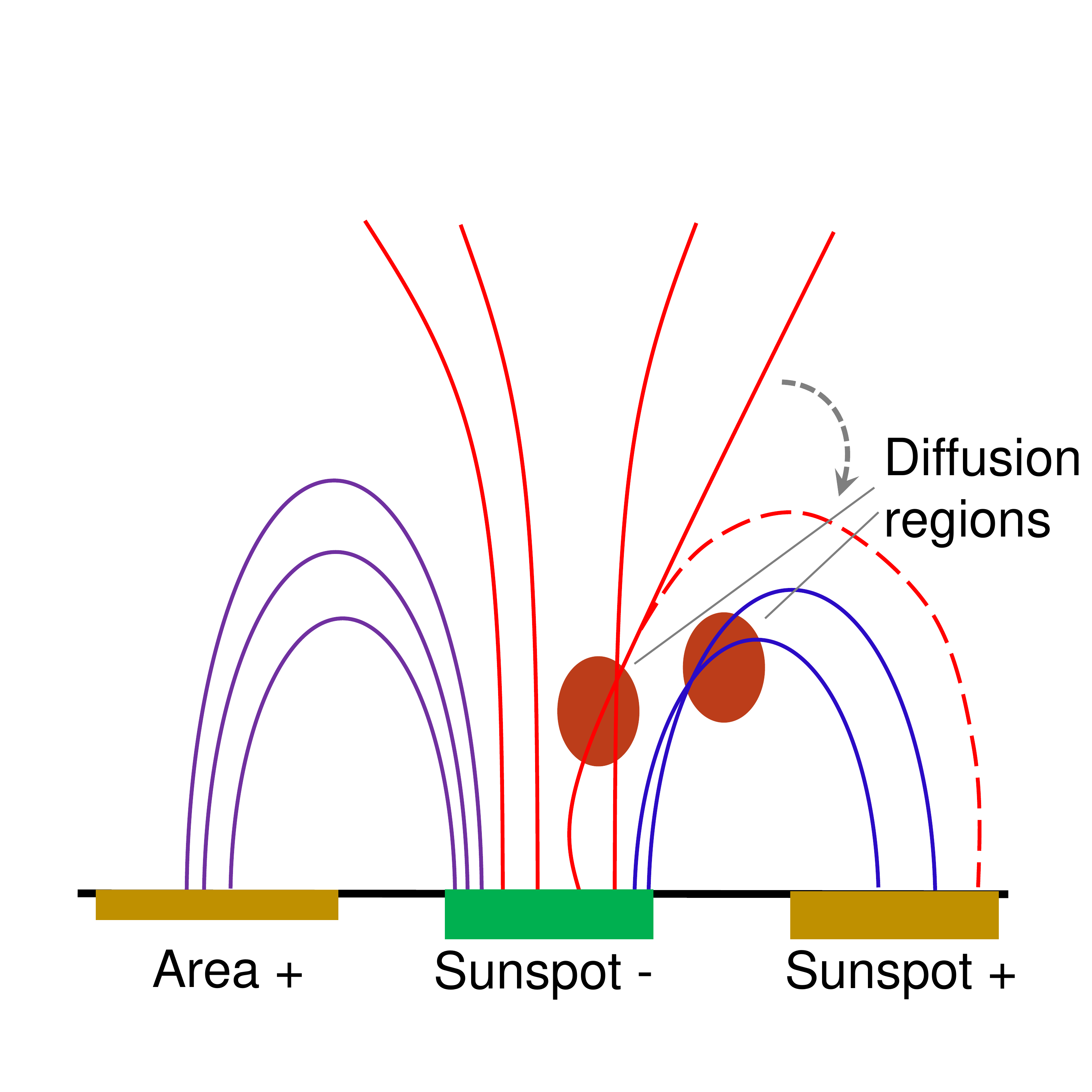}}
\caption{Same as in Figure~4 but for Flare~2.}
\label{f-flare3sketch}
\end{figure}

The reconnection and current dissipation in this configuration result in a change of connectivity. Firstly, there is some apparent field redistribution in the structure C. More interestingly, some open magnetic field from flux~B becomes connected to the negative sunspot, i.e. becomes part of flux~C (Figure~\ref{f-flare3sketch}). Unlike in the Flare~1, this flare does not show any signs of eruption: high velocities are observed only in the immediate vicinity of the diffusion region, corresponding to the reconnection outflow.

There is an important similarity between the two considered events: in both of them, reconnection occurs in close vicinity of a surface separating open and closed magnetic field. Furthermore, in both events, magnetic reconnection results in change of connectivity between open and closed magnetic flux, i.e. both flares demonstrate a similar pattern of so-called interchange reconnection. The main difference between the magnetic field dynamics in the flares is that in Flare~1 some closed magnetic flux becomes open, while in Flare~2 the open magnetic flux fraction remains nearly constant (Figure~\ref{f-fluxratio}). At the same time, the fraction of open magnetic flux going through the reconnection region is decreasing, however, this happens mainly because of the variation of the location and shape of the diffusion region. This has significant implications for accelerated particle escape through the top boundary of the domain, into the upper corona, towards the heliosphere.

\subsection{Particle acceleration and transport}
\label{s-models-particles}

\mrkred{The electric field used in the test-particle part of our model is calculated as $\vec{E} = - \vec{v} \times \vec{B} + \mu_0 \eta_1 \vec{j}$, where $\vec{v}$ and $\vec{B}$ are the local plasma velocity and magnetic field strength, respectively. The background component of diffusivity $\eta_0$, although necessary and unavoidable in MHD simulations, is ignored in particle simulations to prevent unrealistic bulk acceleration. Hence, the typical electric field strength in the diffusion regions in our models can be estimated as $\mu_0 \eta_1 j_\mathrm{crit}$. The adopted value of diffusivity is about 8$\times$10$^9$~m$^2$~s$^{-1}$, while the critical current density is about $j_\mathrm{crit}=1.5\times$10$^{-4}$~A~m$^{-2}$ (see Section~2). Together they yield the value of electric field in the diffusion region of about 1~V~m$^{-1}$. However, during fast Ohmic dissipation the value of electric field in some locations in the diffusion regions reaches about 12~V~m$^{-1}$ in Flare~1 and about 6.2~V~m$^{-1}$ in Flare~2. These values are higher than the value of Dreicer field, which, for typical coronal parameters is about 10$^{-2}$~V~m$^{-1}$ \cite[e.g.][]{tsik06}.

Well outside the diffusion regions, typical current densities are much smaller (by one--two orders of magnitude below the critical current) while the diffusivity $\eta_0$ is 100 times lower than $\eta_1$. Hence, the electric field outside the diffusion region has a non-zero perpendicular component due to plasma advection ($\vec{v}\times \vec{B} \sim 1$~V~m$^{-1}$), while the parallel electric field is practically zero.} 

In both considered events the Ohmic dissipation rate peaks about 50s earlier than the particle acceleration rate. This happens because these two quantities depend on the current density in the diffusion region and on the diffusion region volume in different ways. The Ohmic dissipation rate is proportional to $\eta_1 j^2 \Omega$, where $\Omega$ is the diffusion region volume. At the same time, the number of energetic particles is proportional, approximately, to $\eta_1 j \Omega$, and is sensitive to the shape of the diffusion region. The current density in the diffusion regions is expected to decrease with time, while the diffusion region volumes seem to increase for about 100--150s and then start decreasing. This complex interplay of competing factors is the reason why there is no strong correlation between the energetic particle fluxes and Ohmic dissipation rates.

\begin{figure}
\centering{\includegraphics[width=0.45\textwidth]{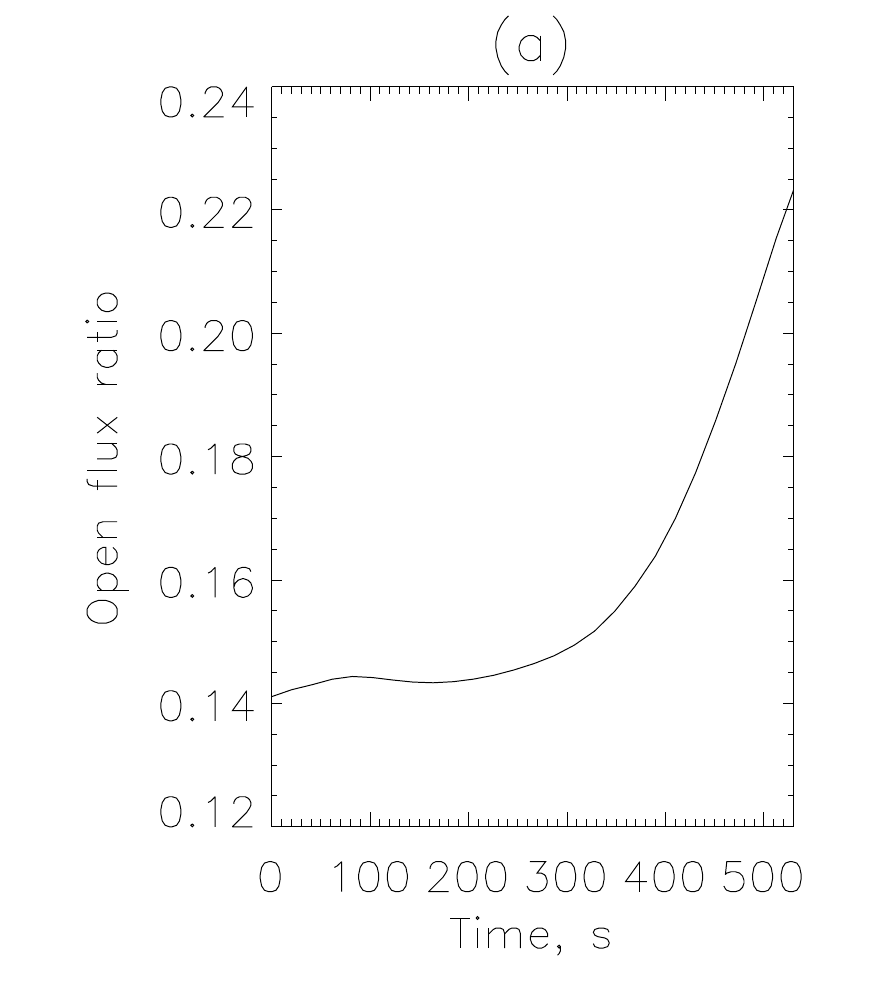}}
\centering{\includegraphics[width=0.45\textwidth]{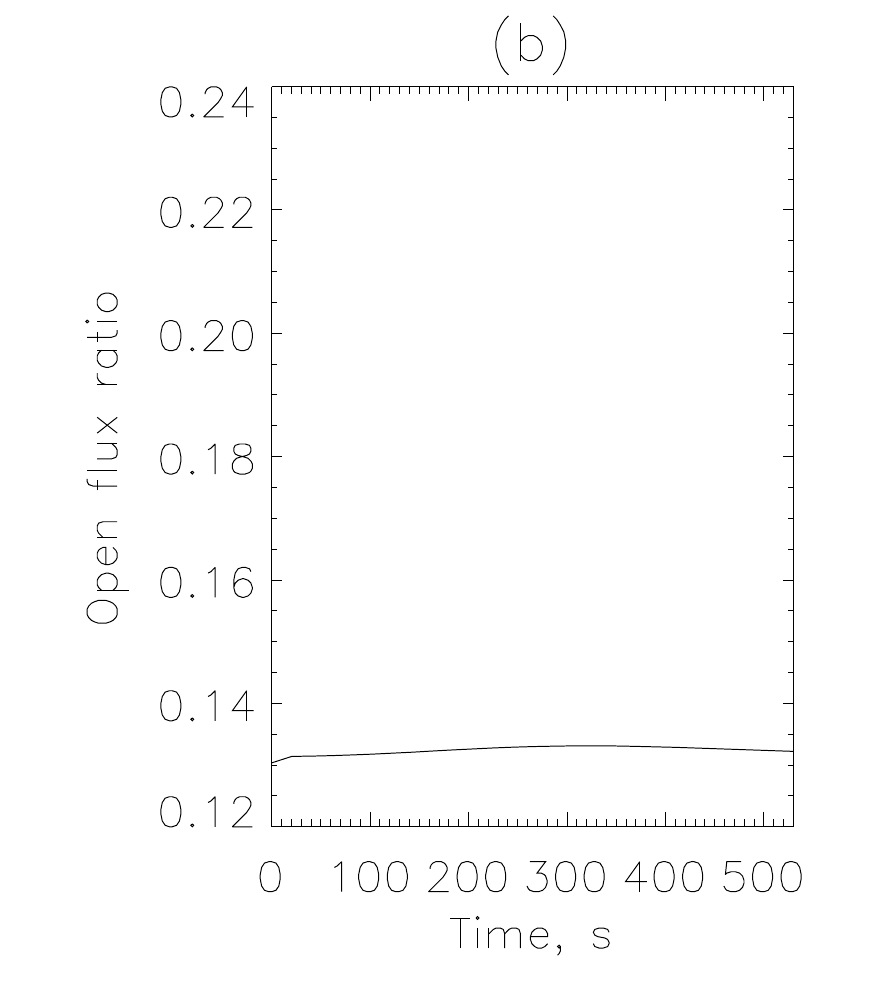}}
\caption{Fraction of the open magnetic flux in the models of Flare~1 (panel a) and Flare~2 (panel b). This fraction is defined as an amount of magnetic flux of the dominant polarity through the top boundary devided by the flux of the same polarity through the bottom boundary.}
\label{f-fluxratio}
\end{figure}

The topology of magnetic field going through the diffusion region is the main factor that determines where energetic particles are transported. In the Flare~1, initially only closed magnetic field goes through the diffusion region (blue lines in Figure~\ref{f-flare1mhd}). However, after only about 25~s, a significant fraction of the field going through the diffusion region is open. This is also true for the model domain as a whole: the fraction of the magnetic flux connecting the lower and upper boundaries of the domain is clearly increasing with time, from 0.14 to about 0.22 (Figure~\ref{f-fluxratio}a).  

In the Flare~2 it is opposite: although initially there is a substantial flux of open magnetic field going through the diffusion region, very quickly (within about 20--30~s) only closed flux remains there. Overall, the fraction of the open magnetic field in the domain remains nearly constant, at about 0.13 (Figure~\ref{f-fluxratio}b).
This does not indicate actual 'closing' of the magnetic field in the domain, it rather indicates the shift of the energy release region towards the closed field.

\begin{figure}
\centering{\includegraphics[width=0.5\textwidth]{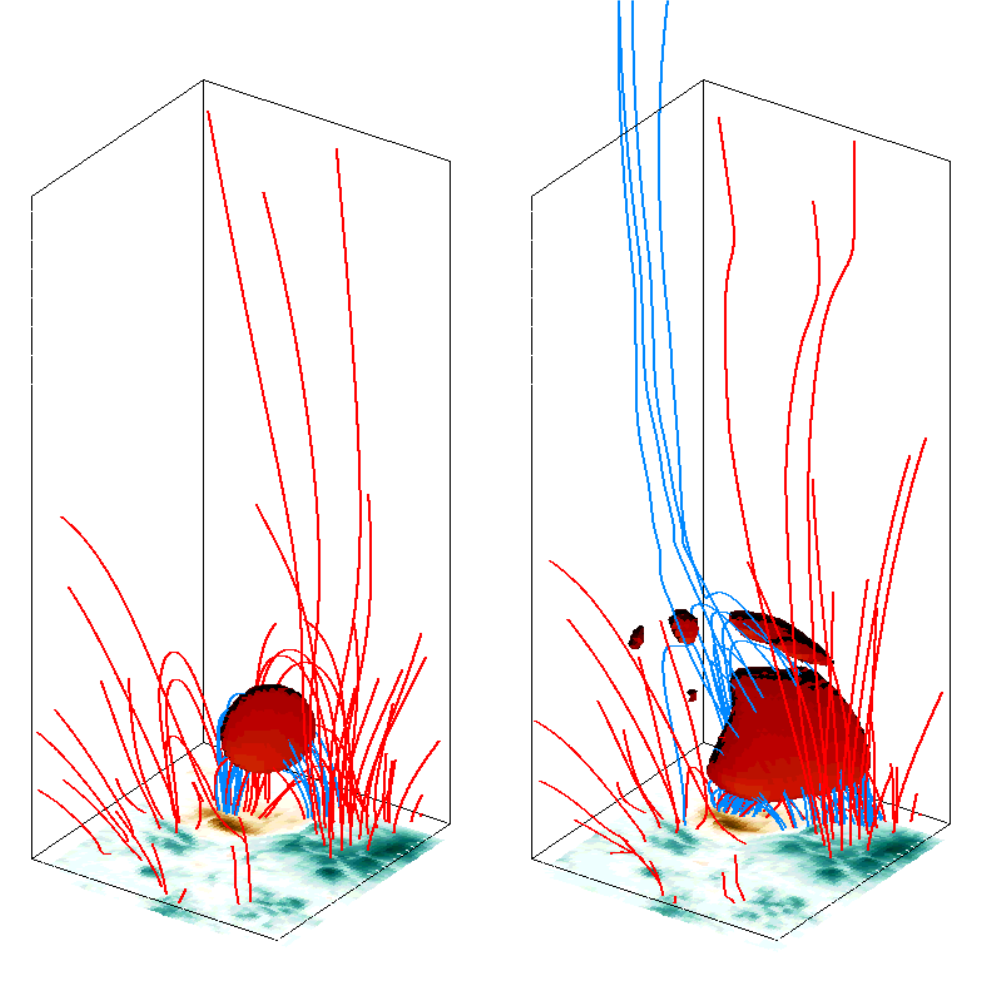}}
\centering{\includegraphics[width=0.5\textwidth]{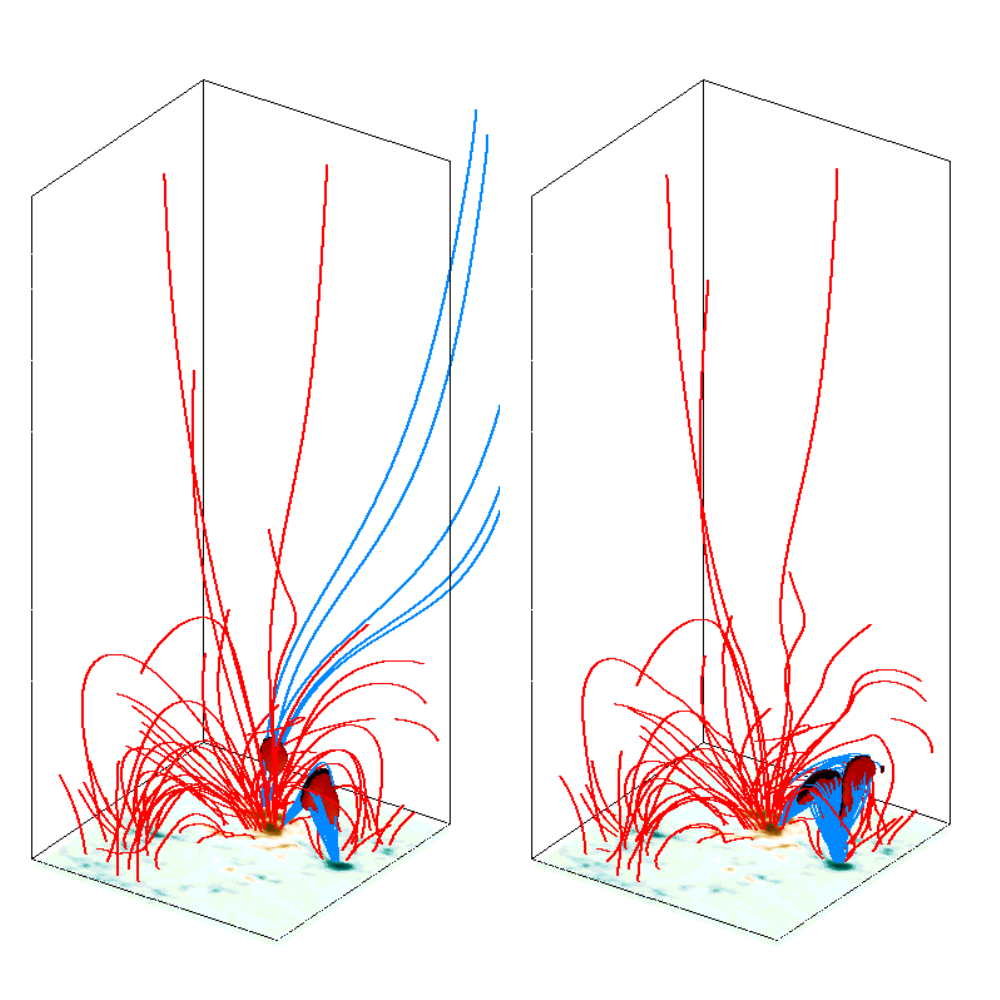}}
\caption{Selected electron trajectories in the model of Flare~1 (panels (a) and (b) showing particles starting at 0 and 72~s, respectively, after onset of reconnection) and Flare~2 (panels (c) and (d) showing particles starting at 0 and 80~s, respectively, after onset of reconnection). Red lines are selected magnetic field lines.}
\label{f-traj}
\end{figure}

Trajectories of selected electrons which become accelerated, superimposed on  the magnetic field, are shown in Figure~\ref{f-traj}. As expected, particles move predominantly along magnetic field lines; however, some drift is visible. This apparent drift consists of the real drift of particles, mainly due to the $E\times B$-drift, representing bulk plasma motion, as well as `effective drift' due to the fact that the magnetic field slightly changes during the travel time of a particle, while it is static at the plot.

\begin{figure}
\centering{\includegraphics[width=0.8\textwidth]{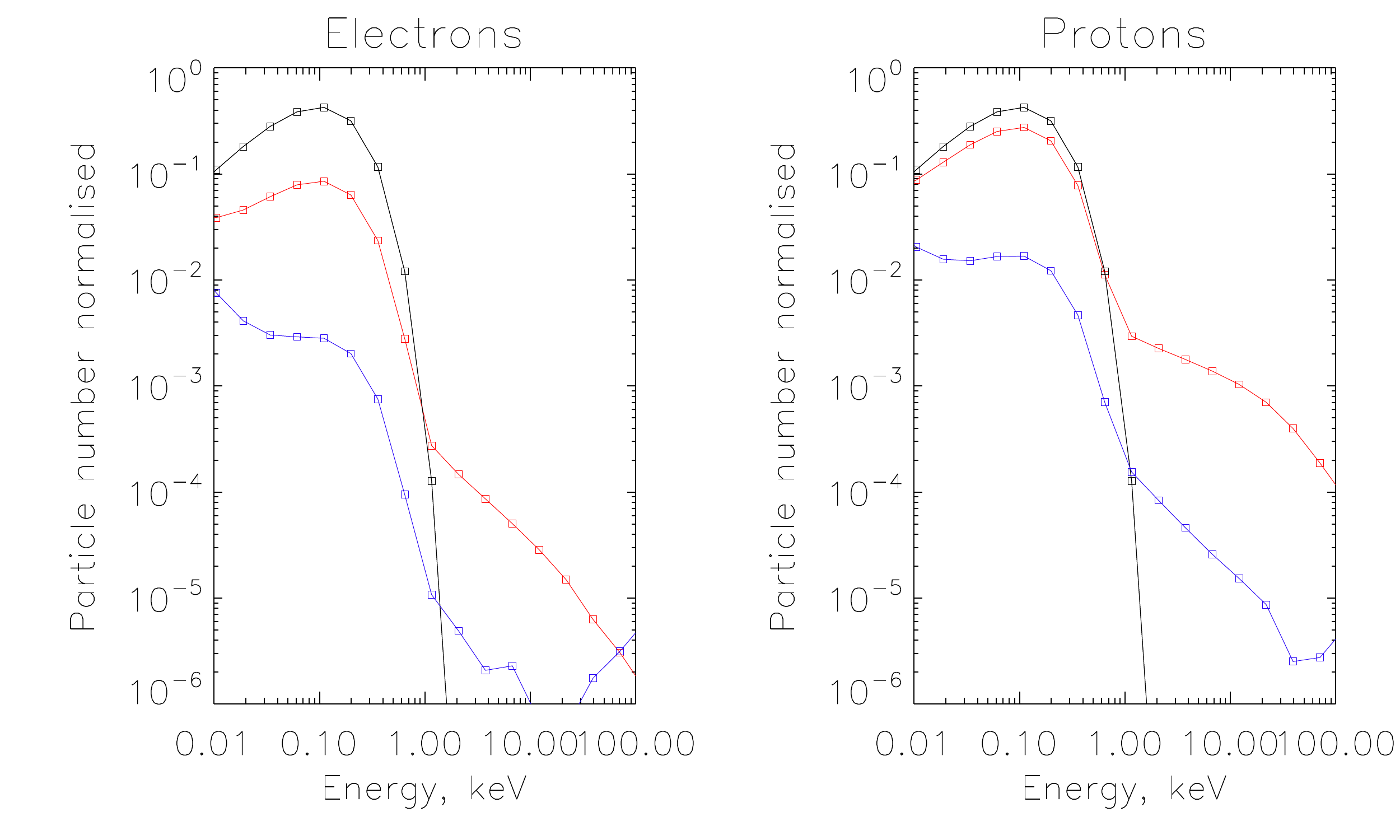}}
\caption{Energy spectra of particles accelerated in Flare~1 during the whole event. Blue and red lines denote spectra for particles which left through the upper and lower boundaries, respectively. Black lines show the initial Maxwellian distributions. }
\label{f-flare1spe}
\end{figure}

\begin{figure}
\centering{\includegraphics[width=0.8\textwidth]{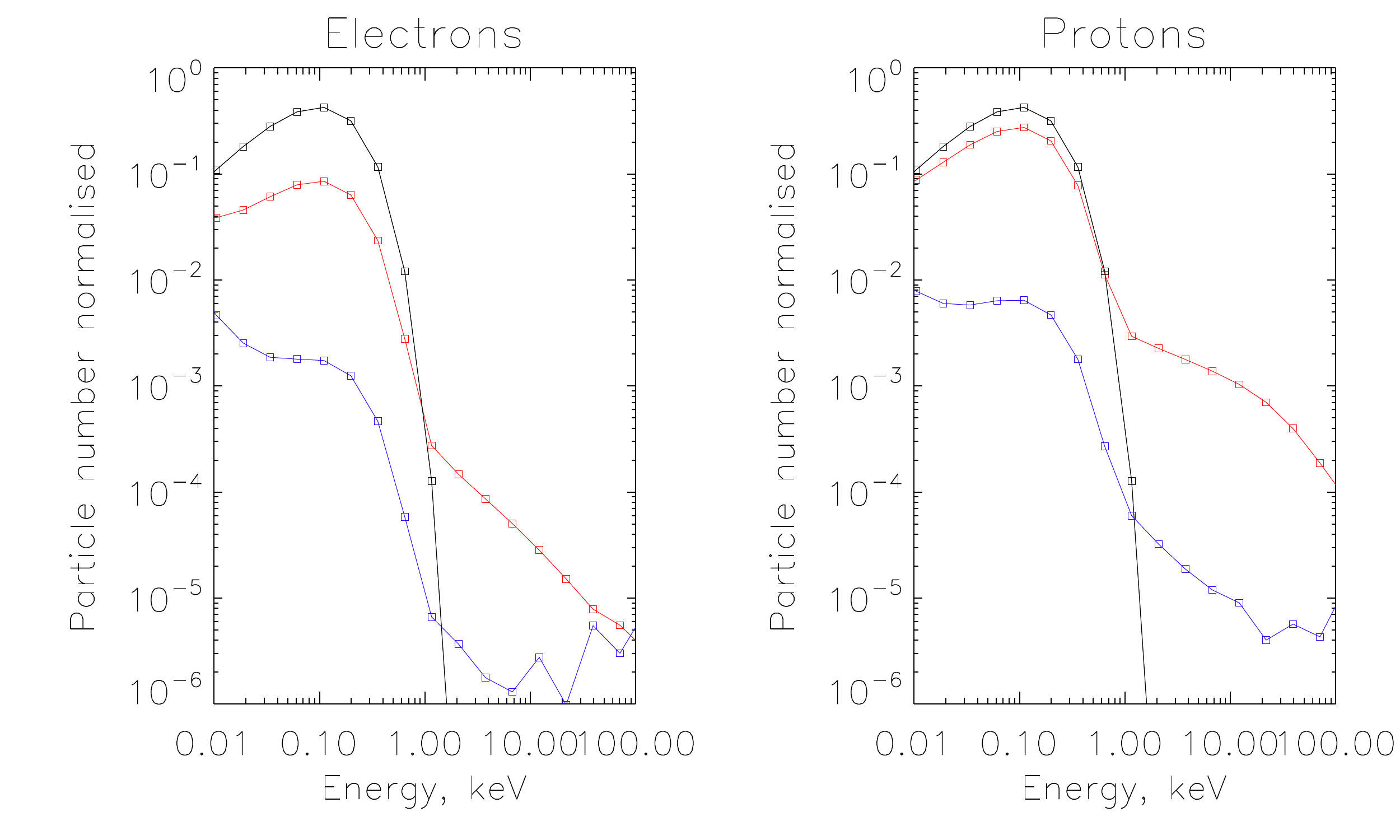}}
\caption{Same as in Figure~10, but for Flare~2.}
\label{f-flare2spe}
\end{figure}

As expected, the majority of test-particles in both models remain thermal: only about 2--3\% of the test-particles in the domain accelerate. The cumulative energy spectra of accelerated electrons and protons are shown in Figures~\ref{f-flare1spe}--\ref{f-flare2spe}. Similar to other test-particle models, the spectra are relatively hard. For the energy spectra  approximated by power-law functions $E^{-\gamma}$ in the range 5--50~keV, the power-law indices $\gamma$ are about 1.5--1.8 for protons and between 1.9--2.5 for electrons. 

Interestingly, the energy spectra of particles leaving through the top and bottom boundaries are different. Both in Flares~1 and 2, the power-law indices of particles going through the top boundary are higher by about 0.5. Thus, in both flare models, the spectral indices of electrons going upwards are 2.4--2.6, while for the precipitating electrons they are about 2.0. Protons reveal a similar pattern: the spectral index $\gamma$ of the population leaving through the top boundary is around 1.8--1.9, while for precipitating protons its is around 1.4--1.5. Hence, our models show that the energy spectra particles moving towards the heliosphere are softer.

The physical mechanism for this difference is not obvious. In both flare models particles get accelerated mainly in the closed field (in flux~B in Flare~1 and in flux~C in Flare~2). Analysis of some particle trajectories shows that some particles leaving through the top boundary get into the open field from the closed field due to cross-field drifts. The dominant component of the cross-field drift, $E\times B$-drift, does not depend on the particle velocity along the magnetic field. Therefore, particles of lower energies, which spend more time close to the contact surface between the open and closed fluxes, will be shifted by $E\times B$-drift to \mrkred{larger} distances. Therefore, lower-energy particles are better at going from one topological magnetic structure to another, and, hence, are more likely to escape from closed magnetic field into open field. This explains why there are more lower-energy particles in the open field, or, in other words, why the particle populations moving towards the heliosphere have softer energy spectra compared to the precipitating particles. 

Variation in the numbers of energetic particles being transported in open and closed fields reflects the dynamics of magnetic field in the flare models and qualitatively can be predicted based on the topology of the magnetic field penetrating the diffusion regions (blue field lines in Figures~\ref{f-flare1mhd} and \ref{f-flare3mhd}). This is particularly true for energetic electrons, which have travel time of few seconds, significantly shorter than the timescale of magnetic field evolution in this model (order of 100~s).

In both events,  particle acceleration starts immediately after the start of evolution in the respective MHD models. The number of electrons leaving the domain in the models of Flare~1 and 2 peaks around 70~s and about 60~s, respectively,  after the start of reconnection. The number of protons leaving the domain peaks about 10-15~s later compared with the electrons, which is consistent with the difference in travel times: energetic protons travel at a speed of about 10$^6$~m/s, which means that their precipitation time from the acceleration sources located about 10--20~Mm above the photosphere is about 10--20~s.

\begin{figure}
\centering{\includegraphics[width=0.8\textwidth]{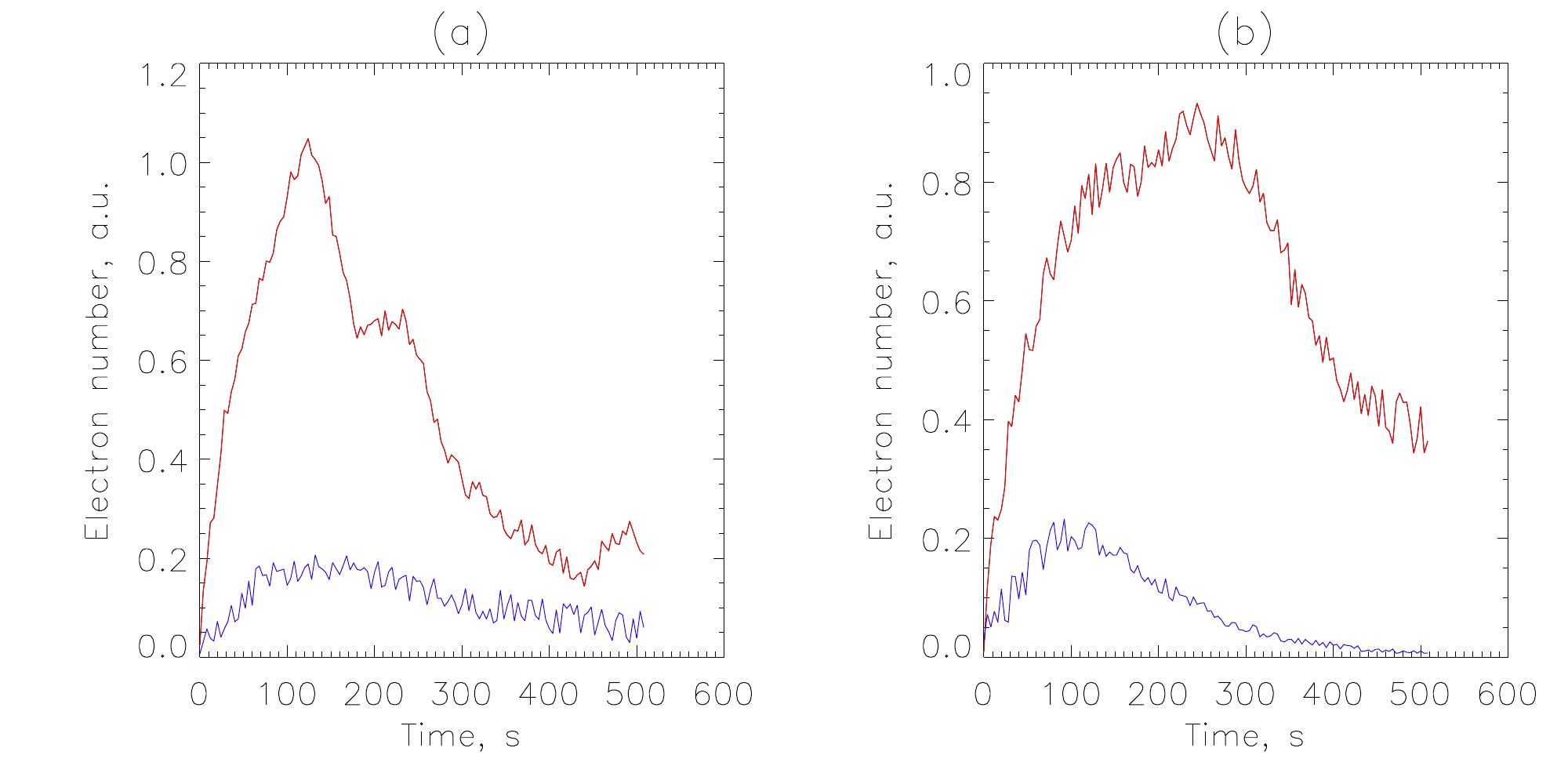}}
\caption{Variation of particle number about 8~keV leaving the domains in Flare~1 and 2, respectively. Blue and red lines show numbers of particles leaving through the upper and lover boundaries, respectively.}
\label{f-flare2supdown}
\end{figure}

More interesting is to compare the energetic particle numbers at the upper and lower boundaries of the domains. At the lower boundary, the flux of energetic electrons and protons above 5~keV approximately follows the Ohmic dissipation rate in both flares. However, the flux of energetic particles through the upper boundary is different (Figure~\ref{f-flare2supdown}). In Flare~1 the fraction of energetic particles leaving through the upper boundary gradually increases for the first $\sim$50~s from zero to about 12--15\% of the total particle number leaving the domain and then remains stable. Overall, about 12-13\% of energetic electrons and protons leave the domain through the upper boundary in this flare. In Flare~2 the situation is different. Initially, about 25--30\% of particles leaving the domain leave through the upper boundary, although there is a substantial uncertainty because of a relatively low number of energetic particles in this model. Within about 100~s the fraction of energetic particles leaving the domain reduces to 1--2\%.

The variation of the fraction of energetic particles leaving through the upper boundary is qualitatively consistent with the amount of open flux through the energy release region. What is interesting, however, is that this fraction is clearly lower than the fraction of the open magnetic flux in the model. Apparently, this happens because the acceleration regions 'favour' closed magnetic field. This is not unexpected: the magnetic field strength in the closed flux should be higher on average than in the open field and the field should contain more free energy.

\subsection{Comparison with observations}
\label{s-models-obs}

The ability of our models to adequately reproduce key features observed in the modelled events is the main criterion for the quality control of the used approach. Based on the obtained simulation results we calculate locations of the precipitation sites for electrons and protons, locations and relative intensities of the bremsstahlung hard X-ray (HXR) sources at the photosphere, and compare them with observations.

\begin{figure}
\centering{\includegraphics[width=0.96\textwidth]{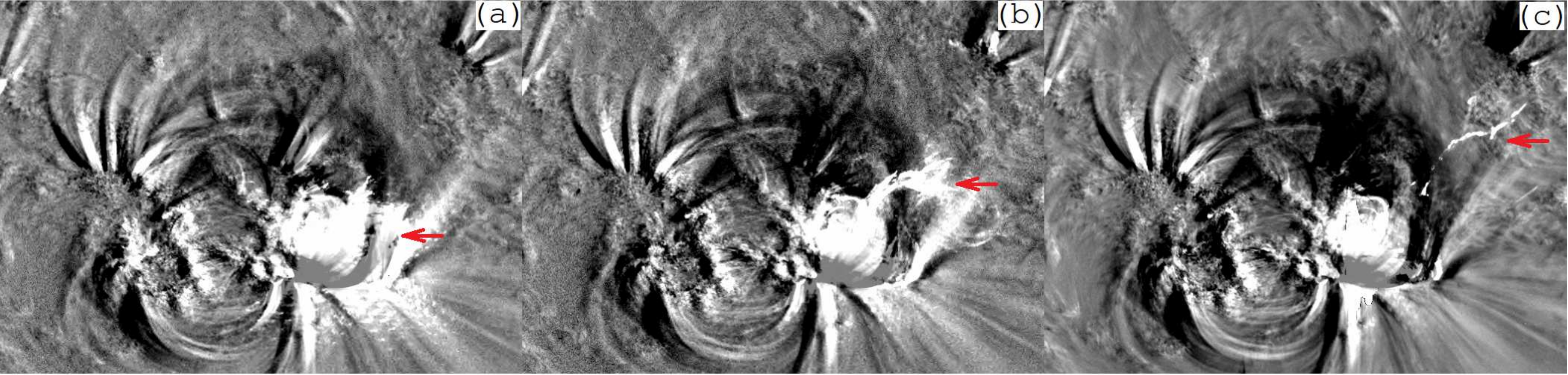}}
\caption{Difference images for 171$\ang$ intensity maps observed by SDO/AIA in Flare~1. Panels (a), (b) and (c) correspond to 22:21, 22:23 and 22:25~UT, i.e. approximately 300, 420 and 540~s after flare onset, respectively.}
\label{f-sdoaia}
\end{figure}

For Flare~1 it is also possible to make a basic comparison between large-scale plasma flows in the model and in the actual event. Figure~\ref{f-sdoaia} shows the difference image for Flare~1 based on the 171$\ang$ intensity maps observed by SDO/AIA. Although there is no large-scale eruption, there is a clear ejection of plasma in the western direction originating approximately at the magnetic loop connecting two strongest sunspots in the active region. This ejection appears to be very similar to the plasma ejection observed in the model (Figure~\ref{f-flare1mhd}): there is an ejection originating approximately in the same location and moving in the same direction. It is difficult to evaluate the speed because of unknown line-of-sight speed of the feature in the actual event, however, they appear to be similar in order of magnitude, $\sim$0.1-0.5~Mm/s. There is an important difference, though: in the model this plasma ejection appears less than 1~min after onset of magnetic reconnection, while in the actual event it starts about 2--3~min later.

The flux of energetic particles per grid-cell at the lower and upper boundaries is calculating by reducing test-particles on the regular grid used for test-particle simulations, using the particle weighting function of cubic shape with the size of 3 grid-spaces (or approximately 2.1~Mm) in $X$ and $Y$ directions and about 8~s in time. This relatively long integration time is necessary to avoid undersampling. This makes it possible to calculate the hard X-ray intensity emission. We do not compute synthetic HXR intensities for different energies, however, because the energy spectra of accelerated particles in both models seem to remain nearly constant with time; the electron fluxes at the lower boundary are the proxy of HXR intensity at all energies.

\begin{figure}
\centering{\includegraphics[width=0.8\textwidth]{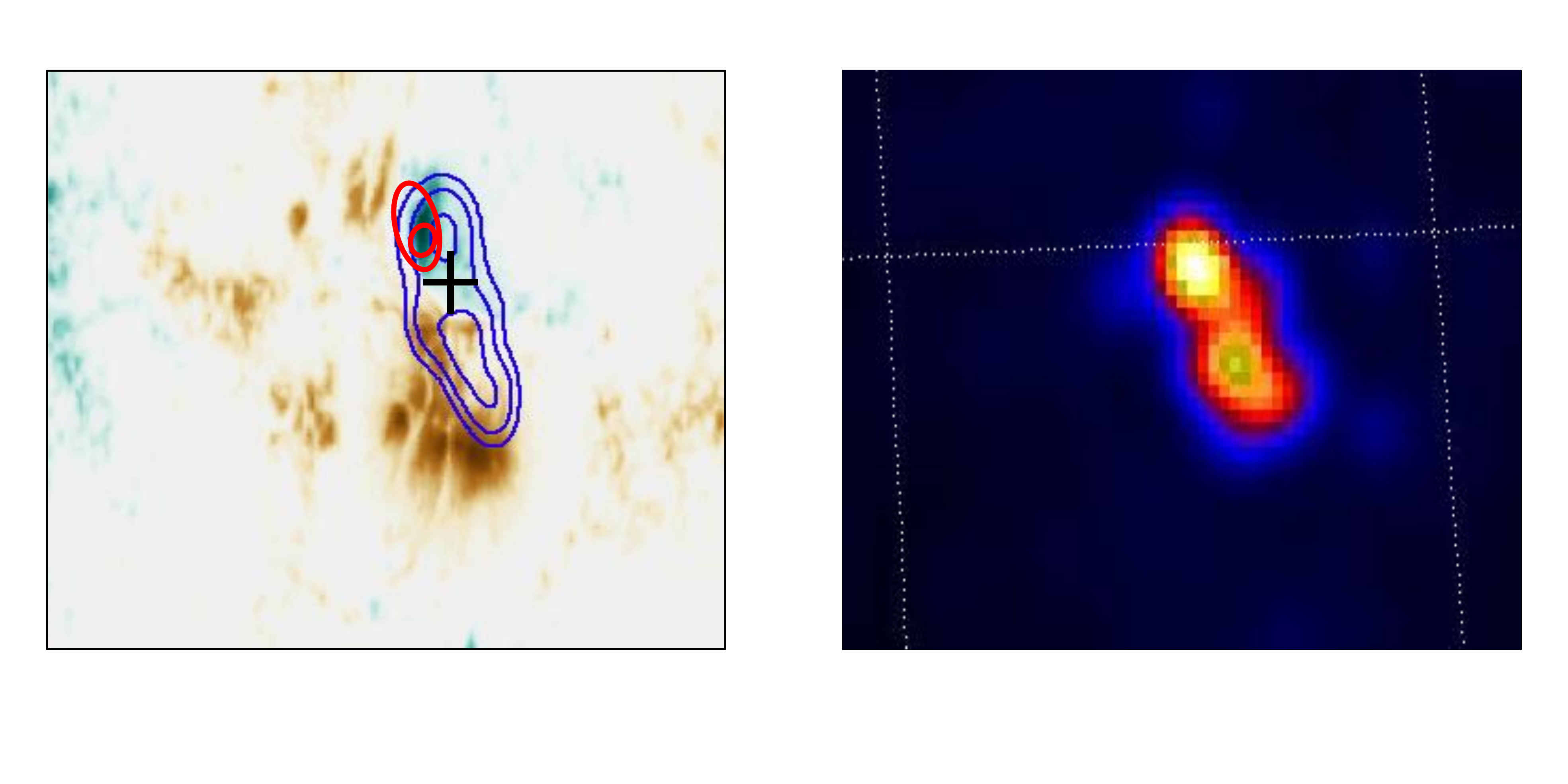}}
\caption{Electron (blue lines) and proton (red lines) precipitation sites in Model~1. Black cross denotes the location of the helioseismic source according to \citet{mace18}. Right panel shows RHESSI intensity map in the 12--25~keV band.}
\label{f-flare1obs}
\end{figure}

\begin{figure}
\centering{\includegraphics[width=0.8\textwidth]{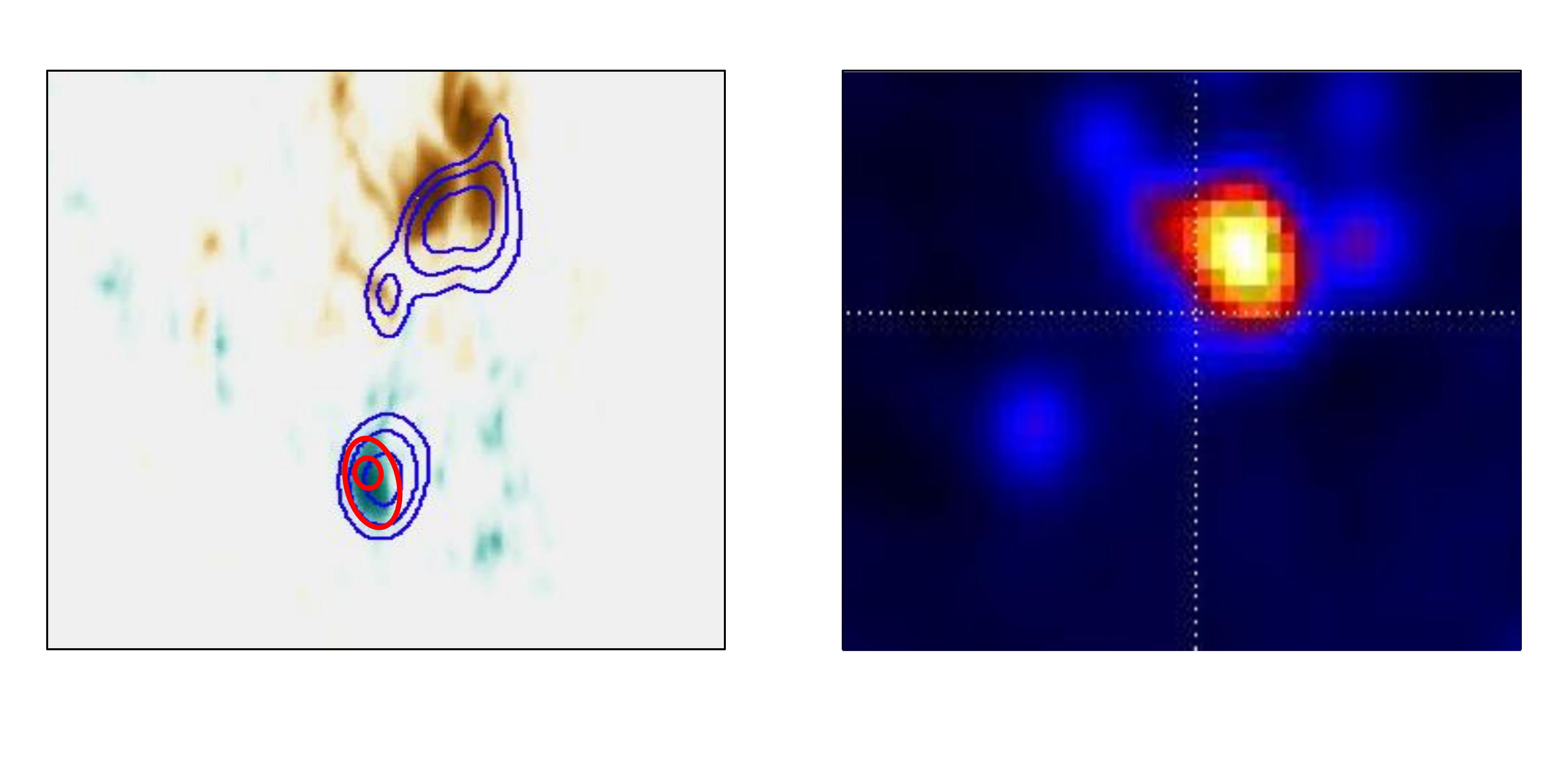}}
\caption{Same as in Figure~13 but for Flare~2.}
\label{f-flare2obs}
\end{figure}

Figures~\ref{f-flare1obs} and \ref{f-flare2obs} show modelled locations of energetic particle precipitation sites compared with HXR observed by RHESSI for Flares~1 and 2, respectively. Additionally,  Figure~\ref{f-flare1obs}  shows the location of the observed helioseismic source in Flare~1. 

It can be seen that in Flare~1 locations, sizes and relative intensities of HXR sources are in good agreement with observations (Figure~\ref{f-flare1obs}). Similar to the RHESSI intensity maps, the synthetic maps clearly show two distinct, although quite close, intensity sources. The brighter source is located over the negative sunspot and has a stretched shape elongated in the south-western direction. Another, slightly fainter source is located over the big positive sunspot and has an elliptical shape oriented in the north-south direction.

The helioseismic source in this flare revealed by \citet{mace18} is located over the big positive sunspots where, according to the particle simulations, most of protons and some of the electrons precipitate. This is similar to the findings by \citet{gore20}, who also find that the helioseismic source appears to be closer to the location where most of energetic protons precipitate, rather than main electron precipitation location. 

Comparing the model of Flare~2 with observations is more difficult because of its relatively low class. This, on its own, is consisten with our modelling, which shows that the diffusion region and the Ohmic heating rate are about 6--7 times lower than in Flare~1. The HXR intensity map observed by RHESSI reveals one large source between 12--25~keV. It seems to coincide with the location of the main electron precipitation site in the model of this flare. There are other fainter sources at this energy as well as a set of spreaded faint sources in the 25--50~keV energy band. However, taking into account relatively low brightness of this flare, it is difficult to say whether these sources are real or artefacts. Furthermore, the shape of the observed bright source is rather different: it is more symmetric than in observations.

It is difficult to compare the duration of the impulsive phase in the flare models with observations. The duration of the impulsive phase in both models flares is about 120--140~s, where by duration we mean the time between the maximum and half-maximum of synthetic HXR emission intensity. The observed duration of the impulsive phase in these events appears to be longer. At the same time, there is a very big uncertainty in measuring these duration: the observed duration appear to be strongly dependent on the X-ray band. 

The impulsive phase in lower-energy emission (below 15--25~keV) has duration of about 250--350~s in both flares. However, the emission at these energies is believed to be produced mainly by thermal electrons \cite[e.g.][]{suie05,case14}. At higher energies, 25--300~keV, which are almost certainly produced by non-thermal electrons, in Flare~1 is about 100--150~s, which is close to the models. In Flare~2 the emission at this energies remains at the background level, i.e. the duration cannot be measured.

\section{Discussion and summary}
\label{s-summary}

As shown in Section~\ref{s-models-obs}, similar to the \citet{gore20} study, the models developed for the considered flares demonstrate good agreement with the spatial structure and relative temporal evolution of hard X-ray. Additionally, the model of the larger event (of 6 September 2011) also demonstrates good agreement with the helioseismic data, as well as the structure and dynamics observed using EUV. This is a significant result in itself: similar to the study by \citet{gore20} we confirm that the MHDTP approach is valid for studying  magnetic field, thermal plasma and energetic particle dynamics in solar flares.

Another similarity with the \citet{gore20} study is that locations of helioseismic sources derived by \citet{mace18} is closer to the locations of when energetic protons, rather than energetic electrons precipitate. Although this is not entirely surprising, because energetic protons transport bigger amount of energy and momentum that energetic electrons (although slower) \cite[e.g.][]{gore05}, this finding contradicts the general consensus that the helioseismic response in solar flares is normally caused by precipitating energetic electrons \cite[e.g.][]{doli05,koso06,zhae20}. However, the separation between the proton and electron precipitation locations is relatively small, meaning there is a substantial uncertainty in linking these locations with the locations of helioseismic sources. Hence, our finding warrants further investigation into this question.

Although the models provide a good match for the relative temporal evolution of the emission intensities, the absolute timing of the event remains uncertain. The initial conditions based on the photospheric magnetograms taken 10--15~min apart result in very similar dynamics of the event. Hence, one can say that MHDTP modelling can be used to explain or predict the structure and dynamics of a solar flare, but not the time of the flare onset.

Furthermore, interestingly, we find that models based on the photospheric magnetograms taken about 30--60~min before the corresponding events provide better agreement with observations compared with models based on the magnetograms taken about 10--20~min before the event. A possible explanation for this effect is that the magnetic field at photospheric level starts restructuring earlier. Evidence of photospheric magnetic field variation starting before flares have been reported in other studies \cite[e.g.][]{wane02}.

Most importantly, our models demonstrate how energetic particles accelerated close to the boundary between the open and closed magnetic field manage to escape through the upper boundary of the model, into the upper corona, towards the heliosphere. Although this contact of two magnetic field topologies plays a crucial role in magnetic reconnection and restructuring, the majority of particles accelerated above 5~keV (more than 80\% both for electrons and protons) gain energy in and are injected into the closed magnetic field. Importantly, the spectral compositions and variations of flux of energetic particles moving in the open and closed magnetic fields are different. Firstly, in both flares particles in the open magnetic field have noticeably harder energy spectra. Secondly, the temporal variation of energetic particle numbers in the open and closed magnetic field are different from those in the closed magnetic fields. In the closed magnetic field energetic particles are present, and, hence, precipitate towards the photosphere during the whole duration of fast energy release phase. Particles moving in the open magnetic field, including those escaping upwards, are present for a shorter time, depending on the dynamics of magnetic reconnection. Thus, in the event which contains some initially closed magnetic field, which becomes open, then number of energetic particles escaping through the upper boundary increases with time, while in the flare where part of initially open flux becomes closed the number of energetic particles escaping upwards decreases much faster than those precipitating towards the photosphere. 

Therefore, in these two flares, although precipitating and escaping particles are accelerated in broadly the same location their number fluxes and energy spectra at the upper and lower boundaries are different. This difference is caused mainly by the structure of magnetic field in and immediately around the reconnection and energy release region. In other words, one can conclude that in the considered events the structure of magnetic field forming the energy release region, on its own, is sufficient to explain the difference between the properties of energetic particle populations precipitating towards the photosphere and escaping towards the heliosphere.

Of course, these conclusions are applicable only to the considered events. Topologically, solar flares are very versatile both, in terms of magnetic field structure and, possibly, in terms of acceleration mechanisms and, hence, we cannot fully extrapolate our findings to other flares. Still, we predict that  interchange reconnection is likely to be responsible for many flares with significant emission of energetic particles into the heliosphere (particularly, in events with no visible eruption). This is because the interchange reconnection is, perhaps, the most probable process in flares involving open field. Therefore, our study offers a viable explanation for the observed differences between energetic particle populations in the solar corona and the heliosphere.

\bibliographystyle{aasjournal}
\bibliography{ngo22}

\clearpage 

\clearpage 

\clearpage 

\clearpage 

\clearpage 

\clearpage 

\clearpage 

\clearpage 

\clearpage 

\clearpage 

\clearpage 

\clearpage 

\clearpage 

\clearpage 

\clearpage 

\clearpage 

\end{document}